\documentstyle[prb,aps,twocolumn]{revtex}

\begin{document}
\draft \flushbottom

\twocolumn[\hsize\textwidth\columnwidth\hsize\csname
@twocolumnfalse\endcsname

\title{Two-domains bulklike Fermi surface of Ag films deposited onto
Si(111)-(7x7)}
\author{J.F. S\'{a}nchez-Royo}
\address{LURE, Centre Universitaire Paris-Sud, Bat. 209 D, B.P. 34,
91898 Orsay Cedex, France\\
and Dpt. F\'{\i}sica Aplicada, ICMUV, Univ. de Valencia, c/Dr.
Moliner 50, 46100 Burjassot, Valencia, Spain}
\author{J. Avila}
\address{LURE, Centre Universitaire Paris-Sud, Bat. 209 D, B.P. 34,
91898 Orsay Cedex, France\\
and Instituto de Ciencia de Materiales de Madrid, CSIC,
Cantoblanco, 28049 Madrid, Spain}
\author{V. P\'{e}rez-Dieste}
\address{LURE, Centre Universitaire Paris-Sud, Bat. 209 D, B.P. 34,
91898 Orsay Cedex, France}
\author{M. De Seta}
\address{Dipartimento di Fisica, Universit\`{a} di Roma III, I-00146
Roma, Italy}
\author{M.C. Asensio\cite{byline}}
\address{LURE, Centre Universitaire Paris-Sud, Bat. 209 D, B.P. 34,
91898 Orsay Cedex, France\\
and Instituto de Ciencia de Materiales de Madrid, CSIC,
Cantoblanco, 28049 Madrid, Spain}

\date{\today}
\maketitle
\begin{abstract}

Thick metallic silver films have been deposited onto Si(111)-(7x7)
substrates at room temperature. Their electronic properties have
been studied by using angle resolved photoelectron spectroscopy
(ARPES). In addition to the electronic band dispersion along the
high-symmetry directions, the Fermi surface topology of the grown
films has been investigated. Using ARPES, the spectral weight
distribution at the Fermi level throughout large portions of the
reciprocal space has been determined at particular perpendicular
electron-momentum values. Systematically, the contours of the
Fermi surface of these films reflected a sixfold symmetry instead
of the threefold symmetry of Ag single crystal. This loss of
symmetry has been attributed to the fact that these films appear
to be composed by two sets of domains rotated 60$^o$ from each
other. Extra, photoemission features at the Fermi level were also
detected, which have been attributed to the presence of surface
states and \textit{sp}-quantum states. The dimensionality of the
Fermi surface of these films has been analyzed studying the
dependence of the Fermi surface contours with the incident photon
energy. The behavior of these contours measured at particular
points along the Ag $\Gamma$L high-symmetry direction puts forward
the three-dimensional character of the electronic structure of the
films investigated.

\end{abstract}

 ]


\section{Introduction}

Silicon (Si) surfaces have been extensively investigated over the
last decades and, consequently, a huge knowledge exists on their
structural and electronic properties.\cite{Haneman,Hasegawa} In
spite of it, some questions still remain open, particularly those
related with their electronic properties and surface dynamics.
\cite{Himpsel3,newphase} Both their fundamental and technological
importance stand for the large interest on the Si surfaces. Among
semiconductor materials, the Si(111) system is one of the most
interesting due to the complexity of its surface structure. The
existence of unoccupied dangling bonds at the bare semiconductor
surface leads to a strong chemical activity on it. In consequence,
the addition of foreign elements can drastically modify its atomic
structure and electronic properties. In this context, a large
interest has been aroused around the reconstructed surfaces
derived by the addition of metallic adatoms, in particular on
first stage of silver (Ag) adsorption, and also on thick metallic
films deposited onto pure Si(111) surface
reconstructions.\cite{Hasegawa}

By far, the most commonly studied Ag/Si(111) systems are those of
Ag films deposited onto the Si(111)-(7x7) and the Ag-derived
Si(111)-($\sqrt{3}$x$\sqrt{3}$)R30$^{o}$ surface reconstruction.
The case of the (7x7) surface reconstruction is specially
attractive since its unit cell shows a reduction in the number of
dangling bonds, compared to the ideal (1x1) surface termination,
and it shows two opposite stacking sequences, one of them having
stacking fault. The peculiar structure of the (7x7) Si termination
determines the dynamics of adatoms at the surface. Scanning
tunneling microscope measurements carried out in submonolayer Ag
films deposited at room temperature onto Si(111)7x7 substrates
showed that Ag preferentially condenses on the faulted halves of
the (7x7) reconstruction.\cite{Ag10ML10,Ag10ML11} In this
situation, films tend to form clusters of few tens of Ag atoms,
already at a coverage of 0.25 monolayers (MLs), with an
interatomic distant close to that of Ag single
crystal.\cite{Ag10ML12,Ag10ML13} As the film thickness increases,
the covered halves of the (7x7) unit cell tend to join each
other,\cite{Ag10ML12} and further deposition tends to cover the
whole surface. After few Ag MLs, multilayered three-dimensional
(3D) Ag islands start to nucleate on the
clusters.\cite{Ag10ML13,Ag10ML14} Impact-collision ion-scattering
spectroscopy measurements revealed that Ag(111) films of several
MLs deposited onto Si(111)-(7x7) at room temperature consist of
domains of Ag(111) crystals rotated 60$^o$ around the surface
normal, in approximately the twin relation,\cite{Ag10ML9} with
their $\langle \overline{1}10\rangle $ directions parallel to
those of the substrate.\cite{Ag10ML8,Ag10ML8b}

In this context, the electronic properties of the surface develop
from those of the (7x7) surface to those determined by metallic
bulklike Ag films. Particularly important are the states closest
to the Fermi level (E$_F$). They are responsible for electron
transport properties as they define the Fermi surface (FS) and a
potential Mott transition at low temperatures. The (7x7) surface
exhibits two nearly dispersionless surface states centered at
binding energies of 0.2 and 0.9 eV below the
E$_F$,\cite{Rowe,Hansson,Houzay,Himpsel1,Uhrberg1,Uhrberg2} which
have been assigned to the dangling-bond and rest-atom states,
respectively. In addition, a back-bond state has been also
observed at 1.8 eV below the E$_F$ near the boundary of the (1x1)
surface Brillouin zone (BZ).\cite{backbond1,backbond2} These states can be
clearly separated from each other and from the bulk semiconductor
electronic states. The possible metallic behavior of this surface
comes through the fact that the surface E$_F$ is strongly pinned
due to the high density of surface states near or at the
E$_F$.\cite{Himpsel2} In order to elucidate the degree of
localization of the characteristic surface states of the (7x7)
surface reconstruction, different techniques have been applied,
such as conductivity,\cite{transporte} nuclear-spin
resonance,\cite{litio} and angle-resolved photoemission
measurements.\cite{Himpsel3} Nevertheless, it is not well
established yet the metallic behavior of the surface state bands
close to the E$_F$. Also, a possible existence of a small gap at
the surface has been discussed.\cite{Himpsel3}

After deposition of Ag adatoms, the electronic properties of the
surface are strongly modified, since they are determined by the
atomic deposition, diffusion at the interface, and nucleation of
the incipient films, which conclude as bulklike thick Ag films. At
first stage of Ag deposition, valence-band spectra photoemission
measurements revealed a quick suppression of the dangling-bond
state of the (7x7) surface with a Ag coverage of 0.2
MLs,\cite{Samsavar} clearly converting the surface into
semiconducting. This result is consistent with scanning tunneling
spectroscopy measurements,\cite{Ag10ML10} which revealed a
semiconducting behavior of the Ag-covered areas of the surface in
comparison to the uncovered ones. Nevertheless, the increase of
resistivity appears to be shorter than that expected for a
suppression of the strongly localized (7x7) surface-state
bands.\cite{Henzler,otroHasegawa,Heike} As the Ag coverage
increases, the surface turns to be metallic, due to the bulklike
behavior of thick Ag films, and its conductivity appears to be
well described by a Drude model, with a linear coefficient close
to the characteristic bulk value at temperatures higher than 40
K.\cite{Ag10ML32,Ag10ML33} Nevertheless, this behavior has not
been connected yet with the electronic properties of the films at
the E$_F$, that is, with the topology of the FS.

In this work, we study the electronic properties and,
particularly, the FS of thick Ag films deposited onto
Si(111)-(7x7) substrates at room temperature, in order to
establish the limiting electronic conditions of the work-frame in
which the conductivity and transport properties of
Ag/Si(111)-(7x7) interfaces can be considered as determined by
metallic Ag bulklike films. For this purpose, angle-resolved
photoelectron spectroscopy (ARPES) appears to be one of the most
powerful and direct tools to probe the electronic structure of
solids. By directly measuring single-particle excitation spectra
(also called spectral weight) as a function of momentum and
energy, it can be determined the most basic quantities of
condensed matter physics, e.g. the band structure, dimensionality
of electronic states, the symmetry and a quantitative analysis of
the FS, electronic gap opening, nesting vector, and so forth.

This paper is organized as follows. Section II summarizes the
experimental details. The results obtained by ARPES are showed and
discussed in Sect. III. In this section it is described the
experimental procedure to obtain by ARPES the spectral function
defining the FS of metallic/semiconductor interfaces. Next, it is
illustrated and discussed the spectral weight at the E$_F$
measured in the studied Ag films, in order to extract and analyze
their FS. We finish this section with the analysis of the
\textsl{quasi}-two-dimensional (\textsl{quasi}-2D) or 3D character
of the investigated interfaces by means of their experimentally
determined FS. The last section (IV) is devoted to summarize the
ARPES results and focus in the conclusions of this work.

\section{Experimental}

   The experiments were performed at LURE (Orsay, France)
using the Spanish-French (PES2) experimental station of the
Super-Aco storage ring, described elsewhere.\cite{Ag10ML1} The
measurements were carried out in a purpose-built ultra-high vacuum
system, with a base pressure of 5x10$^{-11}$ mbar, equipped with
an angle resolving 50 mm hemispherical VSW analyzer coupled on a
goniometer inside the chamber. The manipulator was mounted in a
two-axes goniometer which allows rotation of the sample in the
whole 360$^o$ azimuthal angle and in the 180$^o$ polar emission
angle relative to surface normal ($\Theta _{off}$), with an
overall angular resolution of 0.5$ ^o $. Photoelectrons were
excited with p-polarized synchrotron radiation in the 18-150 eV
energy range. In our experiments, the incident angle of the light
was fixed at 45$^o$. In these conditions, changes of polarization
effects on initial states are neglected.

With this set-up, the procedure to determine the FS using ARPES is
direct. For a given experiment, the photon energy (h$\upsilon $)
was fixed and the intensity at the E$_F$ was recorded, along a
series of azimuthal scans, for each step of the crystal rotation
about its surface normal. This procedure was repeated at different
polar angle positions of the analyzer, which allow us to scan a
sheet of the BZ for each excitation energy. In order to measure
the FS contours at different perpendicular wave vectors, the above
procedure was carried out at different photon energies. In our
measurements, typical polar intervals were 1.5$^o$ and the
azimuthal angle range was fixed at 180$^o$. Depending on
statistics and optimal signal-to-noise ratio, a typical measuring
time of 4 to 6 hours was required to record a FS throughout the
whole BZ.

The p-doped Si(111) single crystal substrate (with a nominal
resistivity of 0.02 $\Omega $cm) was heated up to 650 $^o$C, in
order to degas the wafer, during several hours by resistive
heating and then repeatedly flashed at 1100 C during no more than
15 seconds. Structural order quality was checked out by low-energy
electron diffraction measurements (LEED), which showed sharp spots
corresponding to a distinctive Si(111)-(7x7) surface
reconstruction. In addition to this, no surface impurities were
detected by synchrotron radiation photoemission measurements. The
Si(111)-(7x7) substrate was oriented by azimuthal and polar
photoelectron diffraction (PhD) scans recording the Si 2\textsl{p}
peak intensity.\cite{Ag10ML2} Ag was evaporated onto the surface at
room temperature. The evaporation rate used (of 0.06 MLs/min) was
determined by using a quartz microbalance. In these conditions
thick Ag films, of approximately 10 MLs in thickness, were
deposited. In order to improve their structural quality, the
samples were systematically annealed to 240 $^o$C for 30 min. LEED
spectra measured after growth revealed a clear Ag (1x1) hexagonal
pattern.

\section{Results and Discussion}

\subsection{Determination of the Fermi Surface by using ARPES}

This section is devoted to describe the experimental determination
of the FS of thick Ag films epitaxially grown on Si (111)-(7x7)
substrates by using ARPES. In order to unify the nomenclature to
be used describing our measurements, we show in Figure~\ref{fig1}
the bulk BZ of Ag single crystal together with the characteristic
high-symmetry directions associated to it. The Ag BZ is the
typical one for a fcc material. The (111)-face is hexagonal
alternatively surrounded by hexagonal and square faces. The
schematic representation of Fig.~\ref{fig1} puts forward the
threefold symmetry of the BZ with respect to the [111] direction.
The surface BZ along the [111] direction is hexagonal and its
$\overline{\Gamma }$-point is the projection of the $\Gamma $L
high-symmetry direction of the bulk BZ. The projections of the L
and X points of the BZ lie on the $\overline{M}$ and
$\overline{M'}$ points of the surface BZ, respectively. The
high-symmetry directions of the (111) surface BZ are also
indicated.

Figure~\ref{fig2} shows the well-known theoretical FS of bulk Ag
single crystal oriented as a function of the high-symmetry
directions.\cite{Ascroft} The Ag FS is mainly defined by
\textsl{sp}-states lying at the E$_F$. As predicted by the nearly
free-electron model, the FS is almost spherical with a volume half
of the BZ volume. Nevertheless, the spherical shape of the FS is
distorted. This is a consequence of the hybridization of the
\textsl{sp}-band with the filled \textsl{d} band also close to the
E$_F$, which makes individual spheres touch each other at the BZ
boundary close to the $\left\langle 111\right\rangle $ directions.
Typical \textit{necks} and \textit{bones} are then clearly
identified around the L and X points, respectively.

The FS determined by using ARPES can be easily connected with the
band structure diagrams. Figure~\ref{fig3} shows our results of
calculations of the band structure of fcc Ag single crystal using
a tight-binding hamiltonian. The theoretical simulations allow us
to show clearly the direct relation between the 2D cut of the 3D
FS and the traditional band structure diagrams.
Figures~\ref{fig3}(a) and~\ref{fig3}(b) show the band structure
diagram of Ag single crystal along the $\left[
1\overline{1}0\right] $ and $\left[ 11\overline{2}\right]
$-$\left[ \overline{1}\overline{1}2\right] $ high-symmetry
directions, respectively.\cite{Ag10ML3} In order to calculate
which perpendicular electron-momentum (k$_\perp$) value was proved
when h$\upsilon $=32 eV was used, we have assumed free-electron
final states as well as a work function ($\Phi$) value of
$\Phi$=4.5 eV. With regards to the inner potential (V$_o$) we have
considered it as a parameter. As a consequence of the
no-conservation of k$_\perp$ during the photoemission process, we
find the right value of the V$_o$ by fitting the theoretical
results to the experimental spectral weight at the E$_F$. In this
way, we have obtained a value of V$_o$=-11.5 eV, which is quite
close to that early concluded by Wu \textsl{et al}. (V$_o\sim$-10
eV).\cite{Ag10ML4} The Ag band structure diagrams of
Figs.~\ref{fig3}(a) and~\ref{fig3}(b) show states lying at the
E$_F$ at different parallel momentum (k$_\parallel$) values along
the indicated high-symmetry directions. In essence, the
experimental determination of a 2D cut of the FS by using ARPES
requires scanning of the photoemission signal from states at the
E$_F$ throughout large portions of the reciprocal space. As an
example of the theoretical FS cut in the k$_\parallel$-plane to be
measured by using ARPES, we show in Fig.~\ref{fig3}(c) the 2D FS
cut of Ag single crystal when a photon energy of 32 eV has been
used to perturb the ordered metallic layer.\cite{Ag10ML3}

Photoemission process involves energy and crystal momentum
conservation. In the frame of the three-step model of the
photoexcitation mechanism in ARPES, momentum of the electrons
inside a bulk material can be determined. In this case, both
momentum components, k$_\parallel$ and k$_\perp$, of photoexcited
electrons from states at the E$_F$ can be expressed
as\cite{Ag10ML5}

\begin{mathletters}
\label{ec1}
\begin{equation}
k_{\Vert }=\sqrt{\frac{2m}{\hbar ^2}}\sqrt{h\upsilon -\Phi }\sin
\left(\Theta _{off}\right)\label{mlett:1}
\end{equation}
and
\begin{equation}
k_{\bot }=\sqrt{\frac{2m}{\hbar ^2}}\sqrt{\left( h\upsilon -\Phi
\right) \cos ^2\left( \Theta _{off}\right) -V_o}\;,\label{mlett:2}
\end{equation}
\end{mathletters}

where m is the free-electron mass and $\hbar $ is the reduced
Planck constant. These equations directly imply that, using a
constant h$\upsilon $, all the initial states with a parabolic
shell (k$_\parallel$,k$_\perp$) of the BZ are probed.
Figure~\ref{fig4} shows a transversal cut of the bulk Ag BZ in the
extended zone scheme. This schematic view corresponds to the cut
of the Ag BZ by the plane defined by the $\left[ 111\right] $ and
$\left[ \overline{2}11\right] $ high-symmetry directions. The most
important points are also indicated. The intersection of the Ag FS
of contiguous BZs with that plane is also shown, which forms the
so-called dog-bones contours. In the figure, they are also
indicated successive parabolic shells throughout the bulklike Ag
FS obtained by using Eqs.\ (\ref{ec1}) for h$\upsilon $=32, 55,
and 96 eV. They have been calculated for $\Theta _{off}$ values
ranging between 0$^o$ and 55$^o$, which correspond to our typical
experimental conditions. The intersection between the Ag FS and
these parabolic shells defines 2D FS contours, which pass close to
the middle between the $\Gamma $ and L points for h$\upsilon $=32
eV, near the L point for h$\upsilon $=55 eV, and near to the
$\Gamma $ point for h$\upsilon $=96 eV.

As a summary, we can emphasize that the ability of the ARPES
technique to determine \textsl{in situ} FS contours and band
dispersion of metallic systems, can be envisaged as a very
valuable methodology to experimentally characterize the electronic
properties of novel materials. Nevertheless, it should be pointed
out that extracting the FS contour from the experimental spectral
weight at the E$_F$ can not be considered as an easy task, mostly
in low-dimensional materials,\cite{Mesot} where strong matrix
element variations can occur. In spite of this and, in contrast to
the Haas-van Alphen method, ARPES technique can be used to study a
wide variety of low-dimensional metallic materials like alloys and
highly-correlated electron systems.\cite{Ag10ML6,Ag10ML7}
Additionally, the contours of the FS using ARPES can be measured
throughout all the BZ directions, which allows us to focus a
detailed study of the topology of the FS at particular portions of
the reciprocal space.

\subsection{The Fermi Surface of 10 ML Ag(111) films grown on
Si(111)-(7x7) substrates}

Figure~\ref{holo} shows the experimental 2D FS cut measured in 10
ML-thick Ag films using light of h$\upsilon $=32 eV. The upper
half part of the image corresponds to the measured data. The lower
part is the result of the symmetrization of the measured data with
respect to x-axis. Photoelectron angular distribution measurements
are indicated by taking x-axis as the $\left[
\overline{1}10\right] $ direction of the crystal. The
identification of the high symmetry directions of the surface by
LEED measurements before and after the Ag deposition allows us to
confirm that the Ag(111) films onto Si(111)-(7x7) grow epitaxially
with the Ag overlayer $\langle \overline{1}10\rangle $
high-symmetry directions parallel to those of the the Si
substrate, as it has been previously
reported.\cite{Ag10ML8,Ag10ML8b} The image has been scaled in such
a way that it is linear in photoemission intensity and in
k$_\parallel$. Nevertheless, it should be said that the central
part of the image (k$_\parallel$$<$0.4 \AA $^{-1}$) was
background-enhanced by a constant factor of 2, in order to appear
visible in the image. The photoemission intensity is maximum for
the brightest feature and minimum for the darkest one. In this
image, well-defined features are labelled, indicating the momentum
distribution of initial states lying at the E$_F$ as a function of
k$_\parallel$ and k$_\perp$.

Figure~\ref{fig5}(a) shows the theoretical 2D cut of the ideal Ag
single crystal FS at a particular k$_\perp$ value defined by a
parabolic shell determined by a photon energy of h$\upsilon $=32
eV (see Eqs.\ (\ref{ec1})).\cite{Ag10ML3} This 2D contour reflects
directly the typical threefold symmetry of a fcc material, which
can be described as a distorted ringlike feature (FS1 feature)
with a maximum value of the Fermi momentum (k$_F$) of k$_F$=1.29
\AA $^{-1}$ along the $\langle \overline{1}\overline{1}2\rangle $
directions, a value of k$_F$=1.17 \AA $^{-1}$ along the $\langle
\overline{2}11\rangle $ directions, and a minimum value of
k$_F$=1.13 \AA $^{-1}$ along the $\langle \overline{1}10\rangle $
directions. According to the BZ extended zone scheme, additional
contributions to the FS from neighboring BZs are also expected in
the $\langle \overline{2}11\rangle $ direction (FS2 features) as
well as along the $\langle \overline{1}\overline{1}2\rangle $
direction (FS2' features).

The experimental photoelectron spectral weight image at the E$_F$
(Fig.~\ref{holo}) exhibits most of the characteristic features
predicted for a FS of a bulklike Ag single crystal
(Fig.~\ref{fig5}(a)). Nevertheless, the experimental FS contour
shows a sixfold rather than the typical threefold symmetry of the
FS contour of bulklike fcc Ag single crystal. This clear result is
a consequence of the two-domain character of the Ag films grown
onto Si substrates. Forward scattering PhD has demonstrated that
the metallic films grow epitaxially oriented with respect to the
substrate and that the overlayer consists of two domains of fcc Ag
lattices rotated 60$^o$ from each other.\cite{virginia} As PhD is
a technique that probes and averages large portions of the
interface, we will have two independent contributions to the
photoelectron signal from the two coexisting metallic domains.
Therefore, the photoelectrons with k$_\parallel$ corresponding to
[$\overline{2}11$] and [$\overline{1}\overline{1}2$] directions of
two inequivalent domains will overlap in the recorded
photoemission signal. Fig.~\ref{fig5}(b) shows the theoretical Ag
FS cut calculated for two domains rotated 60$^o$ for h$\upsilon
$=32 eV, which is in excellent agreement with the experimental
data shown in Fig.~\ref{holo}.

In order to carry out a more quantitative analysis of the spectral
weight images as obtained by ARPES we have followed the procedure
proposed by Straub {\sl et al}.\cite{Ag10ML15} The experimental
method used to obtain an image as that shown in Fig.~\ref{holo}
involves selecting an energy window around the E$_F$. The
intensity in this window is represented as a function of
k$_\parallel$ (w(k$_\parallel$) function). In that work, the
authors have analyzed how the width of the energy window, usually
limited by the experimental resolution, affects the measured FS
images. A method was proposed to extract from the FS images the
correct values of k$_F$ as well as the different k$_\parallel$
values of the FS cuts, which can be affected by the experimental
resolution. It involves the calculation of the gradient of the FS
images ($\left| \nabla _{k_\parallel}w(k_\parallel)\right| $).
This procedure produces two maxima out of each maximum of the
spectral weight at the E$_F$ (Fig.~\ref{holo}). It was found that,
out of the two maxima of the gradient, the one on the unoccupied
side of the band crossing the E$_F$ accurately reflects the
k$_\parallel$ values of the FS cuts.

Analyzing the measured spectral weight image by this method, we
have obtained the gradient of the image showed in Fig.~\ref{holo}.
This is displayed in Figure~\ref{fig6}, together with the
two-domains Ag FS cut calculated for h$\upsilon $=32 eV
(Fig~\ref{fig5}(b)). The FS1, FS2, and FS2' features labelled in
Fig.~\ref{fig6} are fairly reproduced by the Ag FS cut calculated.
Nevertheless, the FS2 feature obtained by calculations is shifted
by $\Delta $FS2=0.3 \AA $^{-1}$ respect to the experimental data.
This considerable shift of the contours of the FS mostly appears
in neighboring BZs rather than in the first BZ. This disagreement
between the calculated and experimental Ag FS cut could be
attributed to non-parabolic final states transitions. In fact,
LEED measurements together with constant-final-states
photoemission measurements in the related fcc Cu compound have
demonstrated the influence of unoccupied band structure in the
accuracy of valence band dispersion determination.\cite{Ag10ML18}
Recently, the combined use of LEED and ARPES has revealed that the
unoccupied band dispersion shows a significant deviation from the
free-electron-like behavior. A similar situation could be expected
for the band structure of Ag. The fact that the FS cut calculated
for h$\upsilon $=32 eV shows a large misfit only in neighbors BZs
can also be attributed to a stronger than k$_\perp$-dependency of
the Ag FS as it has been considered in this work (see
Fig.~\ref{fig4}).

Figure~\ref{fig7} shows the profile of both the spectral weight at
the E$_F$ measured with h$\upsilon $=32 eV and its gradient along
the $\langle \overline{1}10\rangle $ directions
(Fig.~\ref{fig7}(a)) and along the $\langle
\overline{1}\overline{1}2\rangle $ and $\langle
\overline{2}11\rangle $ overlapped directions
(Fig.~\ref{fig7}(b)). Following the Straub {\sl et al}.
procedure,\cite{Ag10ML15} we have obtained average values of
k$_F$=1.21 $\pm $ 0.06 \AA $^{-1}$ along the $\langle
\overline{1}10\rangle$ directions and k$_F$=1.34 $\pm $ 0.06 \AA
$^{-1}$ along the $\langle \overline{1}\overline{1}2\rangle $ and
$\langle \overline{2}11\rangle $ directions. These experimental
values are in concordance with those obtained by tight-binding
calculations showed in Fig.~\ref{fig5}.

Whereas the 2D Ag FS cut for h$\upsilon $=32 eV in contiguous BZs
is correctly reproduced by the results of tight-binding, there are
some particular experimental features that are not theoretically
predicted. Figure~\ref{fig6} illustrates the \textsl{s} and
\textsl{h} features. The \textsl{s} feature appears as a rather
diffuse ringlike feature at $\sim $0.26 \AA $^{-1}$ with an
intense spot at the center of the image of 0.10 \AA $^{-1}$ in
radius. The \textsl{h} feature is a diffuse fluted contour at
k$_\parallel$$\sim $0.7 \AA $^{-1}$ that appears under the FS cut
in the first BZ in the $\langle \overline{1}\overline{1}2\rangle $
and $\langle \overline{2}11\rangle $ directions and seems to
connect points of the FS cut in the $\langle
\overline{1}10\rangle$ directions.

In order to analyze the behavior of these unexpected features, we
have measured the band dispersion of the Ag films along the
$\left[ \overline{1}10\right] $ direction and along the $\left[
\overline{1}\overline{1}2\right] $ and $\left[
\overline{2}11\right] $ coincident directions.
Figures~\ref{fig8}(a) and~\ref{fig8}(b) show the spectra obtained
along these high-symmetry directions, respectively. We have
centered our attention in occupied states whose binding energy is
up to 3 eV below the E$_F$. In these figures, we have identified
different photoemission peaks which are indicated by solid bars.
The binding energy of the identified states is plotted versus
k$_\parallel$ for both symmetry directions in Fig.~\ref{fig8}(c)
and~\ref{fig8}(d). Band structure calculations corresponding to
these symmetry directions for h$\upsilon $=32 eV have been also
included in these figures.\cite{Ag10ML3} In the range of binding
energy close to the E$_F$, the band dispersion of the occupied
\textsl{sp} band is the only noticeable structure. In our ARPES
measurements, the \textsl{sp} band dispersion in the first BZ
appears unresolved for both $\left[
\overline{1}\overline{1}2\right] $ and $\left[
\overline{2}11\right] $ directions (Figs.~\ref{fig8}(a)
and~\ref{fig8}(c)). The different states of the \textsl{sp} band
particularly near the E$_F$ are those defining the FS1, FS2, and
FS2' features observed in Fig.~\ref{fig6}. In particular, it
should be noticed that the shift of the FS2 feature of 0.3 \AA
$^{-1}$ at the E$_F$ (Fig.~\ref{fig6}) is also observed in the
occupied \textsl{sp} band up to 3 eV below the E$_F$.

The two occupied states which are not reproduced by the bulklike
Ag band structure are those appearing at normal emission labelled
as \textsl{QW-sp} and \textsl{SS} in Figs.~\ref{fig8}(c)
and~\ref{fig8}(d) with a binding energy of 0.64 and 0.09 eV,
respectively. The presence of the \textsl{QW-sp} state has been
early observed in ARPES measurements in 5-15 ML Ag films onto
Si(111)7x7 substrates.\cite{Ag10ML19} This state has been
identified as a film state associated to the confinement of the
\textsl{sp} band of bulk Ag. The binding energy of confined states
depends on film thickness and on the reflection at the interface,
in the frame of the phase accumulation model.\cite{Ag10ML20} In
our 10 ML Ag films, perpendicular confinement of the \textsl{sp}
states produces confined film-states with a maximum binding energy
round to 0.6 eV according with such a thin film thickness.
Nevertheless, no trace from confined states with higher quantum
number (n) than n=1 has been detected in our measurements. In any
case, this confined state is observed, as expected, to disperse
parabolically through the E$_F$. Out of normal emission, signal
from this state tends to vanish due to the loss of coherence.
Nevertheless, this state gets close enough to the E$_F$ to produce
the ring-like \textsl{s}-feature of $\sim $0.26 \AA $^{-1}$ in
radius observed in Fig.~\ref{fig6}.

Let us now discuss the behavior of the labelled \textsl{SS} state
in Figs.~\ref{fig8}(c) and~\ref{fig8}(d). From band structure
calculations in Ag single crystal,\cite{Ag10ML21} it is not
expected the presence of any bulk state at $\overline{\Gamma }$
point of the surface BZ nor some few eV round the E$_F$. ARPES
measurements carried out in Cu(111), Au(111), and Ag(111) single
crystal show the presence of a parabolic shallow state round to
$\overline{\Gamma }$ point whose dispersion curve is independent
on h$\upsilon $.\cite{Ag10ML22,Ag10ML23,newhufner} This state
crosses the E$_F$ at k$_\parallel$=0.14 \AA $^{-1}$ from
$\overline{\Gamma }$ point in the case of Ag(111).\cite{Ag10ML24}
This state was identified as a Shockley surface state located in
the \textsl{sp} band gap at the L point of the bulk
BZ,\cite{Ag10ML25} whose existence has been attributed to the
break of crystalline periodicity at the surface.\cite{Echenique}
Figures~\ref{fig8}(c) and~\ref{fig8}(d) also reflect the presence
of this surface state \textsl{SS} crossing the E$_F$ at
k$_\parallel$$\sim $0.15 \AA $^{-1}$. Therefore, the central spot
of the \textsl{s} feature observed in Fig.~\ref{fig6} can be
attributed to emission at the E$_F$ from this surface state.

The behavior of the \textsl{h} feature observed in Fig.~\ref{fig6}
is rather different from that observed for the \textsl{s}
features. The cuts of the E$_F$ of the \textsl{sp} band, as shown
in Figs.~\ref{fig8}(c) and~\ref{fig8}(d), appear at the
k$_\parallel$-values observed in the FS cut with h$\upsilon $=32
eV (Fig.~\ref{fig6}). Nevertheless, in Figs.~\ref{fig8}(a)
and~\ref{fig8}(b), no trace of any occupied band appears close to
the E$_F$ in the spectra measured round to 15$^o$-off normal,
corresponding to k$_\parallel$$\sim $0.7 \AA $^{-1}$. In these
spectra, one can only see the trace of a wide tail of the
\textsl{sp} band that lies close to the E$_F$ and whose maximum
appears 4 eV below it (shadow area in Fig.~\ref{fig8}(a)). The
presence of the wide tail of the \textsl{sp} band extending just
below the E$_F$ has been also observed in Ag(111) and in Cu(100)
single crystal by photoemission measurements in normal emission
with h$\upsilon $=7-9 and 21.2 eV and with h$\upsilon $=12-14 eV,
respectively.\cite{Ag10ML27,Goldmannind,Ag10ML28} This fact has
been attributed to indirect transitions induced by the surface.
Moreover, this large tail of the \textsl{sp} band is expected
mostly to reflect the {\sl p}-like behavior of deeper \textsl{sp}
states. Therefore, the presence of the fluted contour in the 2D FS
image of Fig.~\ref{fig6} can be attributed to a consequence of the
surface perturbation process involved in the photoemission
technique.

\subsection{Two- or three-dimensional behavior of the Fermi surface of
the thick Ag(111) films grown on Si substrates}

In order to continue the study on the electronic properties of
thick Ag films grown onto Si(111)-(7x7) substrates we have
centered our further analysis on the 2D or 3D behavior of the FS
of these films. This question has been approached in an early work
by analyzing the \textsl{d} valence band dispersion of Ag films
grown onto Cu(001) single crystal.\cite{Ag10ML29} In that work,
photoemission measurements revealed that the \textsl{d} band shows
already a 3D band dispersion for Ag films of 3-5 ML in thickness.
The radial extent of \textsl{d} states confines their interactions
to nearest neighbors and this interaction is strongly screened by
the \textsl{sp} electrons. Nevertheless, The \textsl{sp} states
extend to additional neighboring sites. These facts would explain
the 3D behavior of the deeper \textsl{d} bands in very thin films,
but they would not necessary imply the same behavior for the FS of
such a thin Ag films. Bulklike behavior of the FS has been also
observed in 1 ML Ni films on Cu(001).\cite{Ag10ML30} In this case,
the 3D behavior of the \textsl{sp} band has been attributed to the
short screening length of electrons in metals and to a strong
hybridization between the Ni \textsl{sp-d}$_{z^2}$ hybrid and the
Cu \textsl{sp}, both of them crossing the E$_F$. Through that
hybridization, the Bloch periodicity of the substrate is imposed
on the electronic wave functions of the film. In Ag films
deposited onto semiconducting Si(111)-(7x7) substrates the
\textsl{sp} states at the E$_F$ have a strong screening effect
inside the substrate. Therefore, the evolution from a 2D to a 3D
Ag FS would come from the fulfillment of periodicity condition of
the \textsl{sp} wave functions inside the films.

The different dimensionality of the metallic films can be clearly
elucidated by using a tuneable light source. Let us consider the
limit case of thin films with 2D electronic properties. In this
case, electronic states have a definite k$_\parallel$ whereas
their k$_\perp$ is undefined, since the overlayer has not enough
periodicity to ensure crystal momentum conservation in the
perpendicular direction. In particular, in the frame of the nearly
free-electron model, the FS of 2D Ag(111) metallic films would be
expected to be ringlike with the ring axis parallel to the $\left[
{111}\right]$ surface vector and contained in the surface BZ. In
this case, FS cuts measured with different incident photon energy
would show FS features at the same k$_\parallel$ values,
independently on h$\upsilon $. Opposite to this, in the case of
thin metallic Ag films with 3D electronic properties both
k$_\parallel$ and k$_\perp$ are defined as indicated in Eqs.\
(\ref{ec1}). Therefore, measurements of the cross-sectional cuts
of the FS with different h$\upsilon $ will indicate a clear
dependence on k$_\perp$. The analytical relation between the
photon energy and the k$_\perp$ allow us to scan the bulklike Ag
FS (Fig.~\ref{fig4}) using different k$_\perp$ values, (i.e. by
changing the photon energy).

Figures~\ref{fig9}(a)-~\ref{fig9}(c) show the theoretical 2D cut
of the Ag single crystal FS at particular k$_\perp$ values probed
with h$\upsilon $=32, 55, and 96 eV, respectively.\cite{Ag10ML3}
The k$_{\parallel x}$ and k$_{\parallel y}$ axes correspond to the
same symmetry directions that those pointed out in
Fig.~\ref{holo}. As expected, all these cuts of the Ag FS put
forward the threefold symmetry of the Ag BZ. As increasing
h$\upsilon $, the scanned reciprocal space increases for a given
experimental angular range. When the photon energy used was
h$\upsilon $=32 eV, the cut of the Ag FS in the first BZ shows the
most relevant features, as described before (Fig.~\ref{fig5}(a)).
For h$\upsilon $=55 and 96 eV cross sectional cuts of the Ag FS
evidence contours from contiguous BZs. For h$\upsilon $=55 eV a
cross section of the \textsl{neck} of the FS at L point could be
probed. As it has been discussed in the previous section, thick
metallic Ag films deposited onto Si(111)-(7x7) substrates are
composed by two domains rotates 60$^o$. Therefore, the FS cuts of
these films, with the assumption of a 3D behavior, would show a
sixfold symmetry rather than the threefold symmetry of the FS cuts
of Ag single crystal for each h$\upsilon $.
Figures~\ref{fig9}(d)-~\ref{fig9}(f) show the cuts of a 3D FS of
Ag films composed of two-domains, calculated at those k$_\perp$
values of Figs.~\ref{fig9}(a)-~\ref{fig9}(c), respectively.

Figures~\ref{fig10}(a)-~\ref{fig10}(c) show the spectral weight
distribution at the E$_F$ measured in a thick Ag film deposited
onto Si(111)-(7x7) with h$\upsilon $=32, 55, and 96 eV,
respectively. These images have been plotted in the same
conditions that those exposed in Fig.~\ref{holo}. All the FS
contours measured at different h$\upsilon $ exhibit the sixfold
symmetry expected for a two-domain Ag film. The FS cut through the
first BZ recorded with h$\upsilon $=32 eV appears as a ringlike
structure (FS1 feature in Fig.~\ref{fig6}). Using photon energies
of h$\upsilon $=55 and 96 eV, the FS cuts through the first
scanned BZ appear as a central bright spot, with a radius of 0.16
$\pm $ 0.08 \AA $^{-1}$, and a ringlike feature, respectively. The
FS cuts in contiguous BZs also show different features as
h$\upsilon $ increases. No trace from surface states can be
distinguished in the h$\upsilon $=55 and 96 eV images. The fact
that the features corresponding to the Ag FS change in
k$_\parallel$-plane with increasing h$\upsilon $ suggests the
presence of bulklike electronic states with a clear 3D character.

To perform a quantitative analysis of the measured FS contours, we
have calculated the gradient of the spectral weight at the E$_F$
images of Fig.~\ref{fig10}, following the procedure suggested by
Straub {\sl et al}.\cite{Ag10ML15}
Figures~\ref{fig11}(a)-~\ref{fig11}(c) show the results of these
calculations for h$\upsilon $=32, 55, and 96 eV, respectively. We
have also included the cuts of the bulklike Ag FS calculated for a
film composed with 60$^o$-rotated domains
(Figs.~\ref{fig9}(d)-~\ref{fig9}(f)). These plots fairly agree
with the experimental results. This good accord between the
calculated and the experimental Ag FS contours improves at higher
photon energy for all the k$_\parallel$-plane scanned, which
suggests that, at high photon energy, the parabolic
nearly-free-electron final-state model could be considered as a
more realistic approximation. These results definitively put
forward the 3D behavior of the FS of 10 ML Ag films deposited onto
Si(111)-(7x7) substrates and, in consequence, also point out that
parameters defining the FS of these Ag films can be compared to
those corresponding to bulklike Ag single crystal.

In order to simultaneously determine k$_F$ and other
characteristic parameters of a 3D FS in different symmetry
directions, it should be taken into account that the measured
k$_F$ values are k$_\perp$-dependent.\cite{AebiSS} Therefore,
reliability of the deduced values depends also on a good k$_\perp$
determination. In our case, the good concordance obtained in the
determination of the Ag FS in the first BZ for all h$\upsilon $
used in this work suggest that k$_\perp$ is reasonably well
determined by a constant inner potential V$_o$=-11.5 eV. In any
case, the influence of V$_o$ on the determination of a bulklike 3D
FS can be overcame with an exhaustive determination of the Ag FS
in a well defined range of h$\upsilon $.

  From our results, on one hand we can directly estimate the
\textsl{neck} radius of the Ag FS at L point from the spectral
weight at the E$_F$ image measured with h$\upsilon $=55 eV, which
appears to be 0.16 $\pm $ 0.08 \AA $^{-1}$. This value of the
\textsl{neck} radius of the Ag FS is in good agreement with those
obtained from other methods.\cite{Ag10ML16,Ag10ML26,Ag10ML31} On
the other side, in the previous section we have estimated the
k$_F$ values for different parallel to the surface symmetry
directions. As The FS of these films appears to show a 3D
behavior, these values of k$_F$ can be also compared with those
obtained in Ag single crystal by the Haas-Van Alphen method
(k$_F$($\Gamma$X)=1.258 \AA $^{-1}$) and those determined by ARPES
measurements in quantum-well \textsl{sp}-states in Ag(100) films
(k$_F$($\Gamma$X)=1.272 $\pm $ 0.002 \AA
$^{-1}$).\cite{Ag10ML16,Ag10ML17} In spite of the small
discrepancy between these two precedent estimates of k$_F$, they
can be considered comparable since the high-precision of the
latter one refers to statistical deviations, without considering
the accuracy of the model used.\cite{Ag10ML17} The fact that our
estimate of k$_F$ appears to be slightly lower than the previous
ones can be attributed to the fact that our measurements do not
exactly probe the FS cut in the $\Gamma$X direction
(Fig.~\ref{fig4}). A simple geometrical calculation
(Fig.~\ref{fig4}) will let us to estimate k$_F$($\Gamma$X)=1.27
$\pm $ 0.06 \AA $^{-1}$, which is in agreement with those
estimates of k$_F$.

Finally, it should be pointed out that the 3D electronic behavior
of the FS of Ag films grown onto Si(111)-(7x7) substrates
establishes the workframe in which the determination of the
transport properties of such a thick films can be carried out. Our
results show that, besides scattering mechanisms involved in
electronic transport, electronic states contributing to current
can be assumed to be very similar to those characteristic of the
bulk metallic material. Therefore, conductivity and transport
properties of these films are expected to be mainly determined by
impurity, phonon, and defectlike scattering mechanisms of
electrons at the E$_F$ with transport parameters close to those of
a single crystal material. In fact, conductivity and
magnetoconductivity measurements carried out in 10 ML Ag films
deposited onto Si(111)-(7x7) substrates at low temperature show a
temperature dependence of resistance well described by a Drude
model, with a linear coefficient at temperatures higher than 40 K
close to that of single crystal.\cite{Ag10ML32,Ag10ML33} These
results appear to contrast with plasmon frequency measurements by
energy loss spectroscopy LEED carried out in up to 18 ML Ag films
deposited onto Si(111)-(7x7) substrates at different
temperatures.\cite{Ag10ML34} In these measurements, plasmon
confinement effects into single domains are attributed to lateral
extent of grains rather than to a perpendicular confinement in the
films, in spite of the metallic conductivity behavior observed in
those films. Mean free path of electrons in those films is round
to some tens of \AA $^{-1}$, at low temperature,\cite{Ag10ML32}
which is of the order of lateral extension of the
grains.\cite{Ag10ML34} These facts suggest that strong lateral
confinement effects are expected mostly to appear in thinner
films, as revealed by magnetoconductivity
measurements,\cite{Ag10ML33} since they show a higher granular
density. In this context, AC conductivity measurements or
high-magnetic field conductivity measurements at low temperature
rather than DC conductivity measurements in 10 films could be
carried out in order to observe lateral confinement effects in 10
ML Ag films. In any case, the fact that films appear to be
composed by two 60$^o$-rotated domains may smooth grain boundary
scattering mechanisms due to strong hybridization of \textsl{sp}
states at grain boundaries. This fact would favor that mean free
path of electrons is mainly determined by film thickness rather
than grain size effects of such a thin films.

\section{Summary AND CONCLUSIONS}

Thick Ag films have been deposited onto Si(111)7x7 substrates at
room temperature. The electronic properties of these films have
been studied by measurements of spectral weight at the E$_F$ at
different h$\upsilon $. With this technique, cross-sectional cuts
of the FS of these films have been obtained for particular values
of k$_\perp$. The FS cuts measured in these films reflect clear
features with a sixfold symmetry. In order to analyze this result,
we have compared the experimental FS cuts with theoretical FS cuts
of 3D Ag single crystal. The calculated Ag FS cuts fairly
reproduces most of the features observed in the experimental data,
but the sixfold symmetry of the Ag films FS is not reproduced.
These facts suggest that the loss of symmetry order of these Ag
films is due to their structural composition. In fact, the Ag FS
calculated for films composed with two domains rotated 60$^o$
fairly reproduces the experimental FS cuts of these films in the
whole space of the BZ analyzed.

In addition to this, photoemission traces at the E$_F$ not
belonging to bulklike Ag FS have been also detected in the FS of
these thick Ag films. They have been attributed to the presence of
surface states, which are located in the \textsl{sp} band gap at
the L point of the BZ, and to \textsl{sp}-quantum states of these
films. Also, additional features have been detected coming from
the tail of deep \textsl{sp} states.

The fact that the FS contours of thick Ag films are well
reproduced by bulklike Ag FS cut for a particular k$_\perp$ value
encouraged us to determine the dimensionality of the FS of these
films. In order to solve this point, FS mapping was carried out by
measuring Ag FS cuts with different photon energies. The FS
contours measured at h$\upsilon $=32, 55, and 96 eV reflect
different traces of the FS as h$\upsilon $ increases, which puts
forward the 3D behavior of the FS of these films. The bulklike Ag
FS cuts calculated for films composed by two-domains, with
k$_\perp$ values corresponding to h$\upsilon $=32, 55, and 96 eV,
fit the FS cuts of these thick Ag films in the whole BZ scanned.

\acknowledgments

This work was financed by DGICYT (Spain) (Grant No. PB-97-1199)
and the Large Scale Facilities program of the EU to LURE.
Financial support from the Comunidad Aut\'{o}noma de Madrid
(Project No. 07N/0042/98) is also acknowledged. J.F.S.-R.
acknowledges financial support from the Ministerio de
Educaci\'{o}n y Cultura of Spain.


\begin{figure}
\caption{Bulk Brillouin zone of the fcc Ag single crystal. Its
projection on the (111)-plane, that is the surface Brillouin zone,
is also drawn. The main points and high-symmetry directions of
both the bulk and the surface Brillouin zones are also indicated.}
\label{fig1}
\end{figure}

\begin{figure}
\caption{Fermi surface of bulk Ag. To clarify the illustration the
bulk Ag Brillouin zone is also drawn, together with its main
points and high symmetry directions.} \label{fig2}
\end{figure}

\begin{figure}
\caption{Bulk Ag band structure calculated by using a
tight-binding method for h$\upsilon $=32 eV in: (a) the $\left[
\overline{1}10\right] $ and (b) in the $\left[
\overline{1}\overline{1}2\right] $ and $\left[
\overline{2}11\right] $ high-symmetry directions. The binding
energy of the bands is referred to the E$_F$, whose position is
indicated in dotted lines. States lying at the E$_F$ are labelled
from E1 to E5. (c) Corresponding Ag FS cut represented in the
k$_\parallel$-plane. Points of the FS cut labelled as E1-E5
correspond to states lying at the E$_F$ indicated in (a) and (b).}
\label{fig3}
\end{figure}

\begin{figure}
\caption{Intersection of final-state nearly free electron
parabolas for h$\upsilon $=32, 55, and 96 eV with the typical
polygons of the Ag Brillouin zone in two selected directions. The
so-called dog-bones structures of the FS cross-section in the
selected plane are also shown. Notice that, by changing h$\upsilon
$, the whole three-dimensional Brillouin zone can be studied.}
\label{fig4}
\end{figure}

\begin{figure}
\caption{(color) Spectral weight at the E$_F$ in a region of the
k$_\parallel$-plane measured in a thick Ag film deposited onto
Si(111)7x7 with h$\upsilon $=32 eV.} \label{holo}
\end{figure}

\begin{figure}
\caption{(a) Fermi surface cut of Ag single crystal represented in
the k$_\parallel$-plane as calculated by using a tight-binding
method for h$\upsilon $=32 eV. The different features obtained are
labelled as FS1, FS2, and FS2'. The Fermi vector is also
indicated. The corresponding Ag surface Brillouin zone is plotted
at the bottom of this figure, in which symmetry directions of the
parallel wave vector are indicated. (b) The same that (a)
calculated for a situation of a film composed with two
60$^o$-rotated domains. At the bottom of this figure, it is
illustrated the fact that the surface Brillouin zones of both
domains appear to be overlapped in ARPES.} \label{fig5}
\end{figure}

\begin{figure}
\caption{(color) Gradient of the spectral weight at the E$_F$
shown in Fig.~\ref{holo}. The Fermi surface cut calculated for
h$\upsilon $=32 eV in the case of a Ag film with 60$^o$-rotated
domains is also plotted in dotted lines. The different features
observed and the Fermi vector are indicated. The misfit in the
position of the FS2 feature is also indicated ($\Delta $FS2).}
\label{fig6}
\end{figure}

\begin{figure}
\caption{Spectral weight at the E$_F$ measured in: (a) the
$\langle \overline{1}10\rangle $ and (b) the $\langle
\overline{1}\overline{1}2\rangle $ and $\langle
\overline{2}11\rangle $ overlapped directions (solid curves). The
gradient of each profile is also plotted under the corresponding
one as dotted curves. The value of k$_F$ extracted from these
gradients is drawn, as average value. Its position is marked by a
solid vertical bar.} \label{fig7}
\end{figure}

\begin{figure}
\caption{Energy distribution curves measured in a thick Ag film
with h$\upsilon $=32 eV along: (a) the overlapped $\left[
\overline{1}\overline{1}2\right] $ and $\left[
\overline{2}11\right] $ and (b) the $\left[ \overline{1}10\right]
$ symmetry directions. The binding energy of these curves is
referred to the E$_F$. The value of $\Theta _{off}$ is given for
some of the curves, being the rest of them equally $\Theta
_{off}$-separated between two consecutive labelled ones. The
different peaks identified have been marked by small solid bars. A
dashed area in the $\Theta _{off}$=14$^o$ curve has been labelled
as {\sl sp-tail}. (c) and (d) (circles, full circles,triangles)
Band dispersion diagram extracted from the peaks identified in (a)
and (b), respectively. (crosses,plus,diamonds) Band dispersion
calculated with h$\upsilon $=32 eV along the $\left[
\overline{1}\overline{1}2\right] $, $\left[ \overline{2}11\right]
$, and $\left[ \overline{1}10\right] $ symmetry directions,
respectively. The E$_F$ position is marked by dotted lines. Bulk
states of the calculated bands lying at the E$_F$ are labelled as
in Fig.~\ref{fig5}(a).} \label{fig8}
\end{figure}

\begin{figure}
\caption{On the left side, Fermi surface cuts in the
k$_\parallel$-plane of Ag(111) single crystal as calculated by a
tight-binding method for h$\upsilon $= 32, 55, and 96 eV. On the
right side, These Fermi surface cuts calculated for a situation of
Ag films composed by two 60$^o$-rotated Ag domains.} \label{fig9}
\end{figure}

\begin{figure}
\caption{(color) (a)-(c) Spectral weight at the E$_F$ in a region
of the k$_\parallel$-plane measured in a thick Ag film deposited
onto Si(111)-(7x7) with h$\upsilon $= 32, 55, and 96 eV,
respectively.} \label{fig10}
\end{figure}

\begin{figure}
\caption{(color) (a)-(c) Gradient of the spectral weight images at
the E$_F$ shown in Fig.~\ref{fig10}, which were obtained with
h$\upsilon $=32, 55, and 96 eV, respectively. The corresponding
calculated Fermi surface cuts in the case of a Ag film with two
60$^o$-rotated domains are also plotted in dotted lines.}
\label{fig11}
\end{figure}


\begin{references}

\bibitem[*]{byline} Electronic address: asensio@lure.u-psud.fr\\
and mcasensio@icmm.csic.es

\bibitem{Haneman}  D. Haneman, Rep. Prog. Phys. \textbf{50}, 1045 (1987).

\bibitem{Hasegawa}  S. Hasegawa, X. Tong, S. Takeda, N. Sato, and T.
Nagao, Prog. in Surf. Sci. \textbf{60}, 89 (1999).

\bibitem{Himpsel3} R. Losio, K.N. Altmann, and F.J. Himpsel, \prb
\textbf{61}, 10845 (2000).

\bibitem{newphase} Y. Fukaya and Y. Shigeta, \prl \textbf{85}, 5150
(2000).

\bibitem{Ag10ML10} St. Tosch and H. Neddermeyer, \prl \textbf{61},
349 (1988).

\bibitem{Ag10ML11} A. Shibata, Y. Kimura, and K. Takayanagi, Surf. Sci.
\textbf{303}, 161 (1994).

\bibitem{Ag10ML12} H. Hirayama, H. Okamoto, and K. Takayanagi, \prb
\textbf{60}, 14260 (1999).

\bibitem{Ag10ML13} K.J. Wan, X.F. Lin, and J. Nogami, \prb
\textbf{47}, 13700 (1993).

\bibitem{Ag10ML14} Z.H. Zhang, S. Hasegawa, and S. Ino, \prb \textbf{55},
9983 (1997).

\bibitem{Ag10ML9} K. Sumitomo, K. Tanaka, Y. Izawa, I. Katayama, F. Shoji,
K. Oura, and T. Hanawa, Appl. Surf. Sci. \textbf{41-42}, 112
(1989).

\bibitem{Ag10ML8} Y. Gotoh, S. Ino, and H. Komatsu, J. Crystal Growth
\textbf{56}, 498 (1982).

\bibitem{Ag10ML8b} Y. Gotoh and S. Ino, Thin Solid Films
\textbf{109}, 255 (1983).

\bibitem{Rowe} J.E. Rowe and H. Ibach, \prl \textbf{32}, 421 (1974).

\bibitem{Hansson} G.V. Hansson, R.I.G. Uhrberg, and S.A.
Flodstr\"{o}m, J. Vac. Sci. Technol. \textbf{16}, 1287 (1979).

\bibitem{Houzay} F. Houzay, G.M. Guichar, R. Pinchaux, P. Thiry,
Y. Petroff, and D. Dagneaux, Surf. Sci. \textbf{99}, 28 (1980).

\bibitem{Himpsel1} F.J. Himpsel, D.E. Eastman, P. Heimann, B. Reihl,
C.W. White and D.M. Zehner, \prb \textbf{24}, 1120 (1981).

\bibitem{Uhrberg1} R.I.G. Uhrberg, G.V. Hansson, J.M. Nicholls, P.E.S.
Persson, and S.A. Flodstr\"{o}m, \prb \textbf{31}, 3805 (1985).

\bibitem{Uhrberg2} R.I.G. Uhrberg, T. Kaurila, and Y.-C. Chao, \prb
\textbf{58}, R1730 (1998).

\bibitem{backbond1} R.J. Hamers, R.M. Tromp, and J.E. Demuth, \prl
\textbf{56}, 1972 (1986).

\bibitem{backbond2}R. Wolkow and Ph. Avouris, \prl \textbf{60}, 1049
(1988).

\bibitem{Himpsel2} F.J. Himpsel, G. Hollinger, and R.A. Pollak, \prb
\textbf{28}, 7014 (1983).

\bibitem{transporte} R. Schad, S. Heun, T. Heidenblut, and M. Henzler,
\prb \textbf{45}, 11430 (1992).

\bibitem{litio} D. Fick, R. Veith, H.D. Ebinger, H.J. J\"{a}nsch, C.
Weindel, H. Winnefeld, and J.J. Paggel, \prb \textbf{60}, 8783
(1999).

\bibitem{Samsavar} A. Samsavar, T. Miller, and T.-C. Chiang, \prb
\textbf{42}, 9245 (1990).

\bibitem{Henzler} M. Henzler, \textsl{Surface Physics of Materials I},
Ed. J.M. Blakely, Academic Press (New York 1975) p. 241.

\bibitem{otroHasegawa} Y. Hasegawa, I.-W. Lyo, and P. Avouris,
Surf. Sci. \textbf{357-358}, 32 (1996).

\bibitem{Heike} S. Heike, S. Watanabe, Y. Wada, and T. Hashizume,
\prl \textbf{81}, 890 (1998).

\bibitem{Ag10ML32} M. Henzler, T. L\"{u}er, and A. Burdach, \prb
\textbf{58}, 10046 (1998).

\bibitem{Ag10ML33} M. Henzler, T. L\"{u}er, and J. Heitmann, \prb
\textbf{59}, 2383 (1999).

\bibitem{Ag10ML1}  J. Avila, C. Casado, M.C. Asensio, J.L. P\'{e}rez, M.C
Mu\~noz, And F. Soria, J. Vac. Sci. Technol. A \textbf{13}, 1501
(1995).

\bibitem{Ag10ML2}  A. Mascaraque, J. Avila, C. Teodorescu, M.C. Asensio,
and E.G. Michel, \prb \textbf{55}, R7315 (1997).

\bibitem{Ascroft} N.W. Ashcroft and N.D. Mermin, \textsl{Solid State
Physics}, Saunders College, Philadelphia, PA.

\bibitem{Ag10ML3}  D.A. Papaconstantopoulos, {\sl Handbook of the Band
Structure of Elemental Solids} (Plenum, New York, 1986). The band
diagrams have been calculated with an empirical second-neighbor
tight-binding Hamiltonian. The orthogonal two-center tight-binding
parameters given by Papaconstantopoulos have been used.

\bibitem{Ag10ML4}  S.C. Wu, H. Li, J. Sokolov, J. Quinn, Y.S. Li, and
F. Jona, J. Phys.: Condens. Matter. \textbf{1}, 7471 (1989).

\bibitem{Ag10ML5}  S. H\"{u}fner, {\sl Photoelectron Spectroscopy},
Springer Series in Solid-State Sciences n. 82 (Springer-Verlag,
Berlin Heidelberg, 1995).

\bibitem{Mesot} J. Mesot, A. Kaminski, H.M. Fretwell, M. Randeria,
J. C. Campuzano, H. Ding, M. R. Norman, T. Takeuchi, T. Sato, T.
Yokoya, T. Takahashi, I. Chong, T. Terashima, M. Takano, T.
Mochiku, and K. Kadowaki, Cond-Mat/ 9910430.

\bibitem{Ag10ML6}  P. Aebi, J. Osterwalder, P. Schwaller, L. Schlapbach,
M. Shimoda, T. Mochiku, and K. Kadowaki, \prl \textbf{72}, 2757
(1994).

\bibitem{Ag10ML7}  M. Lindroos, and A. Bansil, \prl \textbf{77},
2985 (1996).

\bibitem{virginia} V. P\'{e}rez-Dieste J.F. S\'{a}nchez-Royo, J.
Avila, and M.C. Asensio, \textsl{unpublished}.

\bibitem{Ag10ML15}  Th. Straub, R. Claessen, P. Steiner, S. H\"{u}fner,
V. Eyert, K. Friemelt, and E. Bucher, \prb \textbf{55}, 13473
(1997).

\bibitem{Ag10ML18}  V.N. Strocov, R. Claessen, G. Nicolay, S. H\"{u}fner,
A. Kimura, A. Harasawa, S. Shin, A. Kakizaki, P.O. Nilsson, H.I.
Starnberg, and P. Blaha, \prl \textbf{81}, 4943 (1998).

\bibitem{Ag10ML19}  A. L. Wachs, A.P. Shapiro, T.C. Hsieh, and
T.-C. Chiang, \prb \textbf{33}, 1460 (1986).

\bibitem{Ag10ML20} N.V. Smith, N.B. Brookes, Y. Chang, and P.D. Johnson,
\prb \textbf{49}, 332 (1994).

\bibitem{Ag10ML21} H. Eckardt, L. Fritsche, and J. Noffke, J. Phys. F
\textbf{14}, 97 (1984).

\bibitem{Ag10ML22} P. Heimann, H. Neddermeyer, and H.F. Roloff, J. Phys. C
\textbf{10}, L17 (1977).

\bibitem{Ag10ML23} S.D. Kevan, \prb \textbf{33}, 4364 (1986).

\bibitem{newhufner} F. Reinert, G. Nicolay, S. Schmidt, D. Ehm,
and S. H\"{u}fner, \prb \textbf{63}, 115415 (2001).

\bibitem{Ag10ML24} B.A. McDougall, T. Balasubramanian, E. Jensen, \prb
\textbf{51}, 13891 (1995).

\bibitem{Ag10ML25} J.G. Nelson, S. Kim, W.J. Gignac, R. S. Williams,
J.G. Tobin, S.W. Robey, and D.A. Shirley, \prb \textbf{32}, 3465
(1985).

\bibitem{Echenique} P.M. Echenique and J.B. Pendry, J. Phys. C
\textbf{11}, 2065 (1978).

\bibitem{Ag10ML27} T. Miller, W.E. McMahon, and T.-C. Chiang, \prl
\textbf{77}, 1167 (1996).

\bibitem{Goldmannind} T. Michalke, A. Gerlach, K. Berge, R.
Matzdorf, and A. Goldmann, \prb \textbf{62}, 10544 (2000).

\bibitem{Ag10ML28} E.D. Hansen, T. Miller, and T.-C. Chiang, \prl
\textbf{78}, 2807 (1997).

\bibitem{Ag10ML29} J.G. Tobin, S.W. Robey, L.E. Klebanoff, and D.A. Shirley,
\prb \textbf{28}, 6169 (1983).

\bibitem{Ag10ML30} G.J. Mankey, K. Subramanian, R.L. Stockbauer,
and R.L. Kurtz, \prl \textbf{78}, 1146 (1997).

\bibitem{AebiSS} P. Aebi, J. Osterwalder, R. Fasel, D.
Naumovi\'{c}, and L. Schlapbach, Surf. Sci. \textbf{307-309}, 917
(1994).

\bibitem{Ag10ML16}  P.T. Coleridge and I.M. Templeton, \prb \textbf{25},
7818 (1982).

\bibitem{Ag10ML31} A.H. McDonald, J.M. Daams, S.H. Vosko, and D.D.
Koelling, \prb \textbf{25}, 713 (1982).

\bibitem{Ag10ML26} G. Fuster, J.M. Tyler, N.E. Brener, J. Callaway,
and D. Bagayoko, \prb \textbf{42}, 7322 (1990).

\bibitem{Ag10ML17}  J.J. Pagel, T. Miller, and T.-C. Chiang,
\prb \textbf{61}, 1804 (2000).

\bibitem{Ag10ML34} F. Moresco, M. Rocca, T. Hildebrandt, and M. Henzler,
\prl \textbf{83}, 2238 (1999).

\end{references}
\end{document}